# A Statistical Learning Based System for Fake Website Detection

Ahmed Abbasi[1], Zhu Zhang[2], and Hsinchun Chen[2]
*University of Wisconsin-Milwaukee[1], University of Arizona[2]*
*abbasi@uwm.edu, zhuzhang@u.arizona.edu, hchen@eller.arizona.edu*

## Abstract

*Existing fake website detection systems are unable to effectively detect fake websites. In this study, we advocate the development of fake website detection systems that employ classification methods grounded in statistical learning theory (SLT). Experimental results reveal that a prototype system developed using SLT-based methods outperforms seven existing fake website detection systems on a test bed encompassing 900 real and fake websites.*

## 1. Introduction

The increased popularity of electronic markets and blogs has attracted opportunists seeking to capitalize on the asymmetric nature of online information exchange [1]. Many forms of fake and deceptive websites have appeared, including spoof and concocted sites [2, 3]. Spoof sites are replicas of real commercial sites, intended to deceive the authentic sites' customers into providing their information. Concocted sites are deceptive websites attempting to appear as unique, legitimate commercial entities. The quantities of such fake websites are rising at alarming rates. Hundreds of new spoof and concocted sites are detected daily [3].

Existing fake website detection systems are unable to effectively identify fake websites [4]. Their ineffectiveness is attributable to several factors, including: (1) lack of sufficient coverage of different categories of fake websites; (2) inclusion of insufficient quantities and types of fraud cues; (3) the use of simplistic classification rules and heuristics that can easily be circumvented; (4) their inability to dynamically adapt to fraudster's innovations. Collectively, these factors have resulted in systems with inadequate detection rates, consequently garnering diminished user trust and confidence in their abilities [5].

## 2. Statistical Learning Theory and Fake Website Detection

In this study we advocate the development of fake website detection systems that utilize methods grounded in statistical learning theory [6]. Statistical Learning Theory (SLT), also known as Vapnik-Chervonenkis theory, is a computational learning theory that attempts to explain the learning process from a statistical point of view [6]. SLT has brought about the emergence of kernel-based learning algorithms such as support vector machines (SVM). Kernel-based algorithms use a kernel function to map input data into a high dimensional feature space where relations between data instances are found. Using the kernel function, the similarity between data points is found and used to perform classification.

SLT based algorithms provide four important benefits which make them highly suitable for fake website detection. Support for these four characteristics is not exclusive to SLT-based learning methods [7]. However, we are unaware of any other classification method that incorporates all four characteristics.

First, they offer the ability to generalize. SLT promotes the "maximum margin" principle which minimizes the classification error of a classifier while maximizing its ability to generalize [7]. Second, SLT-based methods can handle large sets of input features since they transform input data into a kernel matrix comprised of similarity scores between data points [6, 7]. Third, SLT-based methods such as SVM support the use of custom kernel functions that can be tailored specifically towards their application domain. Fake websites often have stylistic similarities with, and even duplicate content from prior fake websites. Encoding this knowledge using flat vectors of variables is difficult, because of the inherently non-linear structure of websites. An appropriate kernel function could be used to better represent such information. Fourth, learning-based classification systems can more readily adapt to changes in fake website developers' strategies as compared to rule-based systems which rely solely on human knowledge [8]. As with other learning-based classifiers, SLT-based classifiers can also update their models by relearning on newer, more up-to-date website collections.

## 3. The AZProtect System

In order to assess the effectiveness of fake website detection systems grounded in SLT, we developed AZProtect. AZProtect uses an extended set of fraud cues in combination with an SVM classifier. The classifier uses a custom kernel function that is tailored to detect concocted and spoof websites. AZProtect incorporates an extended set of fraud cues comprised of 6,000 attributes stemming from five categories: web page text, source code, URLs, images, and website linkage. These attributes were selected using the information gain heuristic, based on their occurrence distribution across 1,000 training websites.

AZProtect uses an SVM classifier equipped with a kernel function that performs classification at the web page level; all pages within a website of interest are independently classified. The aggregate of these page level classification results is used to predict whether a website is real or fake. The system uses a custom linear composite kernel that considers the average and maximum similarity between a web page of interest and web pages belonging to real/fake websites in the training set. Average similarity is intended to capture the overall stylistic similarity between the web page of interest and those appearing in training websites. Max similarity is designed to measure content duplication.

## 4. Evaluation

We evaluated the effectiveness of AZProtect in comparison with existing fake website detection systems. The comparison tools were seven systems which had either performed well in prior testing or not been evaluated [4]. The comparison systems included SpoofGuard, Netcraft, eBay's Account Guard, IE Phishing Filter, FirePhish, EarthLink Toolbar, and Sitehound. The systems were evaluated on a test bed encompassing 200 legitimate and 700 fake websites (350 concocted and 350 spoofs).

AZProtect significantly outperformed all comparison systems, attaining an overall classification accuracy of 92.56%. It had 10%-15% better fake website detection rates than the Netcraft and SpoofGuard systems, while outperforming other systems by as much as 30%-40%. The results support the notion that systems employing SLT-based methods may be more suitable for fake website detection than existing systems.

## 5. Conclusions

In this study we proposed the creation of fake website detection systems that leverage SLT-based methods. SLT-based fake website detection systems are capable of being applied to the concocted and spoof categories of fake websites, using rich sets of fraud cues and custom kernel functions that consider important characteristics of fake websites. Experimental results affirm the idea that SLT-based systems are better equipped to detect fake websites than existing classification and rule-based systems. Given the high costs associated with online fraud, SLT-based systems supporting enhanced fake website detection capabilities constitute an important endeavor.